\documentclass[twocolumn,table,xcdraw]{aastex631}

\usepackage{hyperref}
\usepackage{nameref}
\usepackage{orcidlink}

\usepackage{silence}
\usepackage{url}
\usepackage{booktabs, multirow}

\usepackage{amsmath}
\DeclareFontFamily{U}{mathb}{\hyphenchar\font45}
\DeclareFontShape{U}{mathb}{m}{n}{
      <5> <6> <7> <8> <9> <10> gen * mathb
      <10.95> mathb10 <12> <14.4> <17.28> <20.74> <24.88> mathb12
      }{}
\DeclareSymbolFont{mathb}{U}{mathb}{m}{n}
\DeclareFontSubstitution{U}{mathb}{m}{n}

\makeatletter
\newcommand\RedeclareMathOperator{%
  \@ifstar{\def\rmo@s{m}\rmo@redeclare}{\def\rmo@s{o}\rmo@redeclare}%
}
\newcommand\rmo@redeclare[2]{%
  \begingroup \escapechar\m@ne\xdef\@gtempa{{\string#1}}\endgroup
  \expandafter\@ifundefined\@gtempa
     {\@latex@error{\noexpand#1undefined}\@ehc}%
     \relax
  \expandafter\rmo@declmathop\rmo@s{#1}{#2}}
\newcommand\rmo@declmathop[3]{%
  \DeclareRobustCommand{#2}{\qopname\newmcodes@#1{#3}}%
}
\@onlypreamble\RedeclareMathOperator
\DeclareMathSymbol{\curvearrowtopleft} {\mathrel}{mathb}{"F0}
\DeclareMathSymbol{\curvearrowbotright}{\mathrel}{mathb}{"F4}
\DeclareMathSymbol{\curvearrowtopright}{\mathrel}{mathb}{"F1}

\makeatother

\newcommand{\ee}{\end{equation}}
\newcommand{\be}{\begin{equation}}
\newcommand{\bea}{\begin{eqnarray*}}
\newcommand{\eea}{\end{eqnarray*}}
\newcommand{\ba}{\begin{array}}
\newcommand{\ea}{\end{array}}
\newcommand{\p}{\partial}
\newcommand{\brh}{{\bar H}}

\renewcommand{\d}{{\mathrm{d}}}
\newcommand{\e}{{\mathrm{e}}}
\renewcommand{\i}{{\mathrm{i}}}

\begin{document}


\title{Dynamics of stellar systems with collisions: eigenvalues and eigenfunctions in nearly collisionless limit}

\author{Evgeny V. Polyachenko}
\affiliation{SnT SEDAN, University of Luxembourg, 
29 boulevard JF Kennedy, L-1855 Luxembourg, Luxembourg}
\affiliation{Institute of Astronomy, Russian Academy of Sciences,\\
48 Pyatnitskaya st, Moscow 119017, Russia}
\email{evgeny.polyachenko@uni.lu, epolyach@inasan.ru}

\author{Ilia G. Shukhman}
\affiliation{Institute of Solar-Terrestrial Physics, Russian Academy of Sciences, Siberian Branch \\
P.O. Box 291, Irkutsk 664033, Russia}
\email{shukhman@iszf.irk.ru}


\begin{abstract}

We examine the decay of perturbations in an infinite homogeneous self-gravitating model with a Maxwellian distribution function (DF) when weak collisions are present. In collisionless systems within the stable parameter range, the eigenvalue spectrum consists of a continuous set of real frequencies associated with van Kampen modes, which are singular eigenfunctions of the stellar DF. An initial perturbation in the stellar density and gravitational potential decays exponentially through a superposition of these modes, a phenomenon known as Landau damping. However, the perturbation in the stellar DF does not decay self-similarly; it becomes increasingly oscillatory in velocity space over time, indicating the absence of eigenfunctions corresponding to the Landau damping eigenfrequencies. Consequently, we refer to perturbations undergoing Landau damping as quasi-modes rather than true eigenmodes.
Even rare collisions suppress the formation of steep DF gradients in velocity space. Ng \& Bhattacharjee (2021) demonstrated that introducing collisions eliminates van Kampen modes and transforms Landau quasi-modes into true eigenmodes forming a complete set. As the collision frequency approaches zero, their eigenfrequencies converge to those of the collisionless Landau quasi-modes.
In this study, we investigate the behavior of the eigenfunction of the least-damped aperiodic mode as the collision frequency approaches zero. We derive analytic expressions for the eigenfunction in the resonance region and for the damping rate as a function of collision frequency. Additionally, we employ the standard matrix eigenvalue problem approach to numerically verify our analytical results.

\end{abstract}



\section{Introduction} \label{sec:intro}

Simplifying the Boltzmann kinetic equation equation by discarding the collision term yields the \cite{Vlasov45} equation.  This simplification is often justified by comparing the collision time with the characteristic dynamical times of the system. When the collision time is significantly longer, the Vlasov equation provides a useful approximation. This simplified equation offers considerable advantages, especially in the analytical study of stellar systems -- for instance, in investigating their stability or the evolution of perturbations arising from various external sources.

It also clarifies conceptual aspects of stellar system dynamics, particularly their eigenmode spectrum. The question of mode spectra in collisionless electron plasmas, analogous to stellar systems, was initially addressed by \citet{Lan_46}. Though not explicitly framed as such, Landau's work examined the evolution of small perturbations in electron density, or electric potential, and distribution function (DF). It demonstrated that density/potential perturbations, at large times, are a sum of decaying exponentials (Landau damping). However, the DF does not decay; it develops increasing corrugation in velocity space. Consequently, no DF eigenfunctions correspond to the Landau damping rates.

Later, \cite{vK_55} and \cite{Case59} clarified how Landau damping, discovered in an initial value setting, relates to the eigenmode spectrum. They showed that Landau damping follows from the decay of an initial perturbation expressed as a continuous superposition of van Kampen (vK) eigenmodes over real frequencies $\omega$. The DF $f_\omega(v)$ of the vK modes is singular at $v=c$ and corresponds to the real eigenvalues $\omega$, or the phase velocity $c=\omega/k$, where the DF and potential perturbations are $\propto \exp[-\i(kx - \omega t)]$. Thus, Landau’s exponentially decaying solutions are not true modes, i.e., not self-similar perturbations, but quasi-modes. The relationship between true and Landau quasi-modes in a gravitating system was analyzed in detail in \cite{PSB}.

Completely neglecting collisions is an idealization that can sometimes lead to contradictions or flawed conclusions. A parallel arises in the hydrodynamics of shear flows: reducing the fourth-order viscous Orr-Sommer\-feld equation to the second-order inviscid Rayleigh equation \citep[see, e.g.,][]{SH2012} can lead to impasses when examining perturbations near the stability boundary. To resolve this, one must revert to a viscous approach, at least formally, because viscosity, though absent from the final formulas, is implicit in their derivation. As detailed in \cite{Lin_55}, this situation prompted the Landau-Lin bypass rule for studies of perturbations in collisionless or inviscid systems.

The question of eigenmodes in collisionless systems, such as electron plasmas or stellar systems, exemplifies situations where neglecting collisions leads to paradoxes. Specifically, the lack of DF eigenfunctions corresponding to Landau quasi-modes, and the emergence of vK modes with their singular structure, are puzzling. Although \cite{vK_55}, \cite{Case59}, and subsequent studies on collisionless plasma and inviscid fluids \citep[see, e.g.,][] {Balmforth_Morrison, PolShu2022} clarified the interpretation of these singular modes, their presence in physical settings remains unsettling.

Accounting for any finite, even arbitrarily small collision frequency $\nu$ helps resolve this issue. At first glance, continuity arguments suggest that the eigenmode spectrum in a collisional medium should, as $\nu \to 0$, revert to the real, singular vK spectrum. However, this expectation proves incorrect: a series of works by Ng and co-authors \citep[see][]{Ng1999, Ng2004, Ng2021} show that including collisions in the kinetic equation, even at $\nu \to 0$, eliminates the vK modes and yields a spectrum of damped modes whose eigenfrequencies align with the Landau quasi-modes at $\nu=0$.

This article examines, numerically and analytically, the nature of eigenfunctions in the regime of infrequent collisions, a topic not fully explored in prior works. We aim to clarify the origin of the DF, specifically how the corrugated structure of the DF, observed in the initial collisionless problem, relates to the collision frequency when collisions are considered. We also investigate how the localization and extent of this corrugation depend on the damping rate. Following the approach of Ng and co-authors, our calculations employ a model of a homogeneous, infinite medium using the Jeans swindle. Our numerical calculations are not limited in accuracy, allowing us to approach the collisionless limit $\nu=0$ by reducing the collision frequency to very small values, significantly smaller than those in previous studies.

Section 2 analytically calculates the eigenvalues $c$ using perturbation theory and examines the eigenfunction structure in the limit of infrequent collisions, focusing on the physically relevant least-damped mode. Section 3 validates these results through high-accuracy numerical calculations using a matrix eigenvalue problem solution. Section 4 discusses the findings.

\section{Analytical Expressions}
\label{sec:analytic}

For very rare collisions, one can derive a correction to the collisionless (Landau) eigenvalues $c=c_{\rm L}$ using perturbation theory with respect to the small parameter $\mu$, the dimensionless collision frequency. 

We begin, as in \cite{Ng2021}, with a linearized system consisting of the one-dimensional Boltzmann equation, incorporating a collision term of the \cite{LenBern} form, 
\be
\frac{\p\, \delta f}{\p t}+v\,\frac{\p\, \delta f}{\p x}-\frac{\p\,\delta\Phi}{\p x}\, \frac{\p f_0}{\p v}=\nu\,\frac{\p }{\p v}\,\Bigl(v\,\delta f+\sigma^2\frac{\p\,\delta f}{\p v}\Bigr)\,,
\label{eq:Bol}
\ee
where $\delta f(t,x,v)$ and $\delta\Phi(t,x)$ represent perturbations of the distribution function (DF) and gravitational potential, respectively. These perturbations are related through the Poisson equation:
\be
\frac{\p^{\,2}\delta \Phi}{\p x^2}=4\pi G \int \hspace{-2pt}\d v \, \delta f \,,
\label{eq:Poisson}
\ee
Here, $G$ is the gravitational constant, $\nu$ is the collision frequency, and the unperturbed DF is assumed to be Maxwellian:
\be
f_0(v)=\frac{\rho_0}{(2\pi\sigma^2)^{1/2}}\,\exp\Bigl(-\frac{v^2}{2\sigma^2}\Bigr)\,.
\ee
Considering disturbances of the form
\be
\delta f = f(v)\,\e^{\i (kx-\omega t)}, \quad \delta\Phi =\phi\,
\e^{\i (kx-\omega t)}
\ee
switching to dimensionless variables, and also using the relation $-k^2\phi=4\pi G\int f\,\d v$, we obtain:
\begin{multline}
    \hspace{-4pt}(u - c)\,g(u) - \eta(u)\hspace{-4pt}\int\limits_{-\infty}^\infty \hspace{-4pt}\d u' g(u')  \\
    = \frac{\mu}{2\i}\frac{\d}{\d u}\hspace{-2pt}\left[ 2ug(u) + \frac{\d g(u)}{\d u} \right].
\label{eq:_main}
\end{multline}
The dimensionless variables are defined as follows. The dimensionless velocity ($u$), phase velocity ($c$), and collision frequency ($\mu$) are given by:
\be
u \equiv \frac{v}{\sqrt{2}\,\sigma}\,, \quad
c \equiv \frac{\omega}{\sqrt{2}k\sigma}\,, \quad
\mu \equiv \frac{\nu}{\sqrt{2}k\sigma}\,.
\ee
The dimensionless unperturbed distribution function $g_0(u)$ and its perturbation $g(u)$ are:
\be
g_0(u) \equiv \frac{\sqrt{2}\sigma}{\rho_0} f_0 = \frac{1}{\sqrt{\pi}} \e^{-u^2}  
\,,\quad 
g(u)   \equiv \frac{\sqrt{2}\sigma}{\rho_0} f\,,
\ee
where
\be
\eta(u) \equiv -\frac{\alpha}{2}\frac{\d g_0}{\d u} = \frac{{\alpha}\,u}{\sqrt{\pi}} \,\e^{-u^2}\,,\ \ 
\alpha \equiv \frac{k_J^2}{k^2}=\frac{4\pi G\rho_0}{k^2\sigma^2}\,.
\ee

First, consider the collisionless case where $\mu=0$. Depending on the parameter $\alpha$, specifically whether it is smaller or larger than 1, initial perturbations will either decay or grow exponentially. In terms of eigenvalues, for $\alpha>1$, there exist two discrete modes: an unstable eigenmode $c_+ = \i \gamma$ with $\gamma > 0$, and a stable eigenmode $c_- = -\i \gamma$, along with a set of singular vK modes \citep[see][for details]{PSB}. At $\alpha=1$, the discrete modes vanish, leaving only vK modes for $\alpha<1$. The eigenfunctions of the discrete modes are given by:
\be
g(u) = \frac{\eta(u)}{u-c_\pm}\,.
\label{eq:g_0}
\ee
These eigenfunctions satisfy the normalization condition, which simultaneously represents the collisionless dispersion relation for the eigenvalues:
\be
\int\limits_{-\infty}^\infty \hspace{-4pt}\d u \,g(u) = \frac{\alpha}{\sqrt{\pi}}\int\limits_{-\infty}^\infty\hspace{-4pt}\d u \, \frac{u\,\e^{-u^2}}{u-c_\pm} = 1\,.
\label{eq:DEq}
\ee

Despite the disappearance of the discrete modes, it can be shown that density and potential perturbations decay exponentially for $\alpha<1$, notably with the damping rate given by the least-damped quasi-mode $\gamma_{\rm L}$ that smoothly continues the growth rate $\gamma(\alpha)$ in the region $\alpha>1$ \citep{INT1974, BT2008}. This solution is obtained by applying the Landau-Lin bypass rule \citep{Lan_46, Lin_55}, which modifies the dispersion relation (\ref{eq:DEq}) as follows:
\be
D(c)\equiv 1-\frac{\alpha}{\sqrt{\pi}}\int\limits_\curvearrowbotright \hspace{-2pt}\d u \,\frac{u\,e^{-u^2}}{u-c} = 0\,.
\label{eq:Deq2}
\ee
Here, the symbol ``$\curvearrowbotright$’’ indicates that the integral is taken along a contour in the lower half of the complex $u$ plane, passing below the point $u=c_{\rm L}$. Since this paper focuses on the modification of $c_+$ and $c_{\rm L}=-\i\,\gamma_{\rm L}$ in the presence of collisions, and both belong to the same branch obtainable from relation (\ref{eq:Deq2}), we will use $c_{\rm L}=-\i\,\gamma_{\rm L}$ for all $\alpha$, although the index `L' is somewhat misleading for the unstable mode when $\alpha>1$. Note, however, that no eigenfunction on the real $u$-axis corresponds to the Landau solution for $\alpha<1$. For the latter, it is also convenient to define the parameter
\be
\beta \equiv 1-\alpha
\ee
because $\beta=0$ corresponds to the marginal stability of our model.

\medskip

We now apply to 
$g(u)$ the perturbation theory with respect to $\mu$:
\be
g(\mu, u)=g^{(0)}(u) + g^{(1)}(\mu, u),\ \ c=c_{\rm L}+c_1(\mu)\,.
\ee
Here, $c_{\rm L}$ represents the collisionless `eigenvalue' satisfying (\ref{eq:Deq2}). For the zeroth-order approximation of the eigenfunction $g^{(0)}(u)$, we will use (\ref{eq:g_0}), with $c_\pm$ replaced by $c_{\rm L}$, which yields:
\be
g^{(0)}(u)=\frac{\eta(u)}{u-c_{\rm L}}.
\label{eq:g0}
\ee
The key point here is that, although this is not an eigenfunction of the collisionless problem for $\alpha<1$ on the real $u$-axis, it is the eigenfunction on a complex-valued $u$-contour that passes below $c_{\rm L}$ \citep{PSB}. The perturbed quantities $g^{(1)}(\mu, u)$ and $c_1(\mu)$ are assumed to be of order ${\cal O}(\mu)$. To first order in $\mu$, substituting into (\ref{eq:_main}) gives:
\begin{multline}
- c_1 g^{(0)}(u) + (u - c_{\rm L})\, g^{(1)}(\mu, u)\\
- \frac{\alpha}{\sqrt{\pi}}\,u\, e^{-u^2} \int\limits_\curvearrowbotright \hspace{-2pt}\d u' g^{(1)}(\mu, u') \\
= \frac{\mu}{2\i}\frac{\d}{\d u}\hspace{-2pt}\left[ 2u g^{(0)}(u) +  \frac{\d g^{(0)}(u)}{\d u} \right].
\end{multline}
Dividing the equation by $(u-c_{\rm L})$ and integrating over $u$ along the lower contour yields:
\begin{multline}
\hspace{-2pt}-c_1\hspace{-2pt} \int\limits_\curvearrowbotright \hspace{-2pt} \frac{\d u\, g^{(0)}(u)}{u - c_{\rm L}}
+ \int\limits_\curvearrowbotright\hspace{-2pt}\d u\, g^{(1)}(\mu, u) 
\Bigl(1 - \int\limits_\curvearrowbotright\hspace{-2pt}\d u\, \frac{\alpha}{\sqrt{\pi}}\frac{u\, e^{-u^2}}{u-c_{\rm L}}  \Bigr)\\[2mm]
= \frac{\mu}{2\i} \int\limits_\curvearrowbotright \frac{\d u}{u - c_{\rm L}}\, 
\frac{\d}{\d u} \left[ 2u g^{(0)}(u) +  \frac{\d g^{(0)}(u)}{\d u} \right].    
\end{multline}
The round bracket vanishes due to (\ref{eq:Deq2}). Then, it is straightforward to obtain $c_1 = -\i \mu\, c_{\rm L}\, I_2 /(2I_1)$, where
\bea
I_1 \equiv \frac{\alpha}{\sqrt{\pi}}\int\limits_\curvearrowbotright\hspace{-2pt}\d u\, \frac {u\,\e^{-u^2}}
{(u-c_{\rm L})^2}\,, \ \
I_2\equiv \frac{\alpha}{\sqrt{\pi}}\,\int\limits_\curvearrowbotright\hspace{-2pt}\d u\,
\frac{\e^{-u^2}}{(u-c_{\rm L})^4}\,.
\eea
The explicit expressions for these integrals are (see Appendix A):
\begin{align}
I_1 &= \frac{1}{c_{\rm L}}  \left(\beta-2\,c_{\rm L}^2\right),\\[2mm]
I_2 &= {\frac{4}{3}}+{\frac{2}{3}}\,\left(\beta-2\,c_{\rm L}^2\right)={\frac{4}{3}}+\frac{2}{3}\,c_{\rm L}\,I_1.
\end{align}
Finally, we obtain
\be
\frac{c_1}{\mu}= -{\frac{\i}{3}}\,\left[1+\frac{2}{\beta-2\,c_{\rm L}^2(\alpha)}\right]\,c_{\rm L}^2(\alpha).
\label{eq:final_c_1}
\ee
For the least-damped aperiodic mode of interest, we obtain the first-order correction in $\mu$ to the damping rate of the Landau quasi-mode, $c_1=-\i\,\Delta\,\gamma\equiv -\i\, [\gamma(\mu)-\gamma_{\rm L}]$:
\be 
\gamma' \equiv \lim\limits_{\mu\to 0} \frac{\Delta\gamma}{\mu} = -\frac{1}{3}\,\gamma_{\rm L}^2\,\left(1+\frac{2}{\beta+2\,\gamma_{\rm L}^2}\right).
\label{eq:DGamma}
\ee
Near the stability boundary, where $|\beta| \ll 1$:
\be
c_{\rm L}\approx -\i\,\frac{\beta}{\sqrt{\pi}}\,,\quad
\frac{\Delta\gamma}{\mu} \approx -\frac{2}{3\pi}\,\beta\,.
\label{eq:DGamma0}
\ee
This implies that the damping rate correction changes sign across the stability boundary:  in the stable region where $\beta>0$, it acts to destabilize the system (reducing the damping rate), while in the unstable region where $\beta<0$, it acts to stabilize the system (reducing the growth rate). Exact numerical calculations, which do not rely on perturbation theory in the small parameter $\mu$, confirm this approximate analytical result with high precision and extend it to the second order in $\mu$. Note also, that Equation (\ref{eq:DGamma}) is consistent with the results obtained by \cite{Chavanis2013} using a different approach.

We now seek an approximate analytical description of the least-damped mode eigenfunction $g(\mu, u)$ near resonance. For the aperiodic mode (Re $c=0$) with small damping rates, this eigenfunction is localized near $u=0$. We refer to this narrow region around $u=0$ as the \textit{inner region}, and the rest of the $u$-axis as the \textit{outer region}. We proceed using matched asymptotic expansions \citep[see, e.g.,][]{Malley2014}, a technique for solving singularly perturbed differential equations. We introduce $h(u)$ by writing $g(\mu, u)=h(u)\,\exp(-u^2)$, which transforms the equation (\ref{eq:_main}) to:
\begin{multline}
(u-c)\,h -\frac{\alpha}{\sqrt{\pi}}\, u \hspace{-4pt} \int\limits_{-\infty}^\infty \hspace{-4pt}\d u\, h(u)\,\e^{-u^2} \\ =
\frac{\mu}{2\i} \left( \frac{\d^2 h }{\d u^2} - 2u \frac{\d h }{\d u} \right).
\label{eq:h1}    
\end{multline}
We express the eigenvalue as $c=-\i\gamma(\mu)$ and introduce the inner variable $U\equiv u/\delta$, also defining $\Gamma\equiv \gamma/\delta$. We then set
\be
\delta=({\mu}/{2})^{1/3}.
\ee
It is worth noting that in the hydrodynamic stability theory of nearly inviscid shear flows $U(y)$, a similar spatial scale, $\ell_\nu=\nu^{1/3}$, known as the width of the viscous critical layer, is introduced near the critical level $y=y_c$, where $U(y_c)=c$ and $\nu$ is the inverse of the Reynolds number \citep[see, e.g.,] [] {Benney&Maslowe1975, Chu_Shu1987, ChSh1996, ChurShukh1996}. This scale also arises in weakly collisional plasma in connection with plasma echo \citep{Su_Oberman1968}.

Imposing the same normalization for $g(u)$ as in (\ref{eq:DEq}), we obtain, to leading order in $\delta$, the inner problem equation:
\be
\frac{\d^2 h}{\d U^2} - \i\,(U + \i\Gamma)\,h=-\frac{\i\,\alpha}{\sqrt{\pi}}\,U,
\label{eq:h_ini}
\ee
which must be matched with the inner asymptotics of the outer solution. The outer solution is given by
\be
h = \frac{\alpha}{\sqrt{\pi}}\,\frac{U}{U+i\Gamma} \approx \frac{\alpha}{\sqrt{\pi}}\,\left(1-\i\,\frac{\Gamma}{U}\right)\,.
\ee
Hence, the matching condition requires that $h\to {\alpha}/{\sqrt{\pi}}$ as $|U|\to \infty$.
It is therefore convenient to first extract this constant by setting $h(U)=H(U)+{\alpha}/{\sqrt{\pi}}$. For $H(U)$, we obtain the equation
\be
\frac{\d^2 H}{\d U^2}-\i\,(U+\i\Gamma)\,H=-\frac{\alpha}{\sqrt{\pi}}\,\Gamma,
\label{eq:H}
\ee
with $H(U)$ decreasing as $|U|\to\infty$:
\be
H(U)\to -\i\,\frac{\alpha\,\Gamma}{\sqrt{\pi}\,{U}}\,.
\label{eq:Ha}
\ee
Equation (\ref{eq:H}) is now amenable to solution using the Fourier transform. For 
\be
{\tilde H}(q)=(2\pi)^{-1}\int\limits_{-\infty}^\infty \hspace{-3pt}\d U\, H(U)\,\e^{-\i qU} \,,
\ee
we obtain a first-order equation:
\be
\frac{\d{\tilde H}(q)}{\d q}+(\Gamma-q^2)\,{\tilde H}(q)=-\Gamma\,\frac{\alpha}{\sqrt{\pi}}\,\delta(q),
\label{eq:dif_H_q}
\ee
where $\delta(q)$ is the Dirac delta function. It is straightforward to solve this equation:
\be
{\tilde H}(q)=\frac{\alpha\,\Gamma}{\sqrt{\pi}}\,\exp(q^3/3-\Gamma\,q)\,\Theta(-q),
\ee
where $\Theta(x)$ is the Heaviside step function. After inverse Fourier transformation, we have:
\be
H(U)=\frac{\alpha\,\Gamma}{\sqrt{\pi}}\int\limits_0^{\infty}\hspace{-2pt}\d q \,\exp\left[-\i\,q\,(U+\i\Gamma) - {q^3}/3\right] \,.
\label{eq:H_int}
\ee
It is readily verified by integration by parts in (\ref{eq:H_int}) that for $|U|\gg 1$, the outer limit of the inner solution matches the inner limit of the outer solution (\ref{eq:Ha})
as expected. Finally, for $h(u)$, we obtain:
\be
h(u)=\frac{\alpha\,\gamma}{\delta\,\sqrt{\pi}}\hspace{-2pt} \int\limits_0^{\infty}\hspace{-2pt}\d q\,\exp\left(\frac{\gamma-\i u}\delta \,q 
 -\frac{q^3}3\right) + \frac{\alpha}{\sqrt{\pi}}.
\label{eq:hu}
\ee

\section{Numerical Validation}

For the numerical evaluation of the eigenvalue problem at finite $\mu$, we will adopt an expansion in a series of normalized Hermite polynomials $\brh_n(u)$, similar to \cite{Ng1999}:
\be
g(\mu, u)=\e^{-u^2}\,h(u)=\e^{-u^2}\sum_{n=0}^\infty {A}_n \brh_n(u),  
\label{eq:gu}
\ee
where $\brh_n(u)={H_n}(u)/\bigl({2^{n/2} \sqrt{n!}\,\pi^{1/4}}\bigr)$ and 
\be
H_n(u)=(-1)^n\,\e^{u^2}\, \frac{\d^{\,n} }{\d u^n}(\e^{-u^2})
\ee
are the physicist's Hermite polynomials \citep[see, e.g.,][] {Gradshteyn_8}. For $n=0$, we obtain $\brh_0(u)=\pi^{-1/4}$, so that normalizing $g(\mu, u)$ to unity yields ${A}_0=\pi^{-1/4}$ for all $\mu$.

Equation (\ref{eq:h1}) leads to the following system of linear equations for the expansion coefficients ${A}_n$:
\begin{align}
    {A}_1 &=\sqrt{2}\,c\,{A}_0\,, \\
    {A}_2 &=(c+\i\mu)\,{A}_1-(\beta/\sqrt{2})\,{A}_0\,,\\
    {A}_{n+1} &=\sqrt{\frac{2}{n+1}}\left[(c+\i\mu\,n)\,{A}_n-\sqrt{\frac{n}{2}}\,{A}_{n-1}\right]
\end{align}
for $n\ge 2$. 

However, in contrast to \cite{Ng1999}, we formulate this problem as a standard matrix eigenvalue problem. For the aperiodic mode of interest, where $c(\mu) = -\i\gamma(\mu)$, it is convenient to rewrite it in a purely real form. Introducing ${\tilde\gamma}=\sqrt{2}\,\gamma$, ${\tilde\mu}=\sqrt{2}\,\mu$, and ${A}_n=\i^n\,\tilde A_n$, we obtain:
\be
{\tilde\gamma}\,\tilde A_n=\sum\limits_{m=0}^\infty {\cal M}_{nm}({\tilde\mu})\,\tilde A_m,  \label{eq:matrix_eq}
\ee
where ${\cal M}_{nm}(\tilde \mu)$ is a tridiagonal matrix:
\begin{align*}
{\cal M}_{nm}({\tilde\mu})\!=\!\left|
\ba{cccccc}
0&-\sqrt{1}&0&0&0&...\\
\beta &\tilde\mu&-\sqrt{2}&0&0&...\\
0&\sqrt{2}&2\tilde\mu&-\sqrt{3}&0&...\\
0&0&\sqrt{3}&3\tilde\mu&-\sqrt{4}&...\\
0&0&0&\sqrt{4}&4\tilde\mu&...\\
.&.&.&.&.&...\\
\ea\right| \,.
\end{align*}
The eigenvector $\tilde A_n$ then contains only real values.

The sum in the matrix equation (\ref{eq:matrix_eq}) is truncated at a large number $n_{\rm max}$. To solve this, we employed the Matlab sparse matrix function \texttt{eigs}, enhanced by the Advanpix package for multiple precision calculations. This function enables us to find a single eigenvalue close to a pre-set approximate value for $n_{\rm max} \simeq 10^7$ with quadrupole precision (128 bits) on a workstation with 256 GB of memory. Our experiments indicate that this is sufficient to reliably determine the solution for $\mu\gtrsim 10^{-7}$.

For smaller values of $\mu$, we implemented finding eigenvalues and eigenvectors in C from scratch, utilizing the GNU Multiple Precision Floating-Point Reliable Library. This allowed us to increase $n_{\rm max}$ to $10^9$ with quadrupole precision (128 bits), enabling us to reach $\mu \lesssim 10^{-9}$. The characteristic equation for determining $\tilde\gamma$ is given by the vanishing of the determinant of the truncated matrix ${\cal M}_{nm}(\tilde \mu)-\tilde\gamma\, \delta_{nm}$, i.e.,
\be
    \tilde\gamma\, P(\tilde\mu, \tilde\gamma) + \beta\, Q(\tilde\mu, \tilde\gamma) = 0\,,
\label{eq:ev_trunc}
\ee
where $P$ and $Q$ are polynomials in $\tilde\gamma$ and $\tilde\mu$ of order $n_{\rm max}$. This code calculates an eigenvalue from Eq. (\ref{eq:ev_trunc}) for a given parameter $\mu$ using the bisection method. The corresponding eigenvector is then approximated using a method that combines inverse iteration with a direct solve for a tridiagonal system. 

\begin{figure} [b]
\centering
\includegraphics[width=87mm, trim={7pt 9pt 0pt 0pt}, clip]{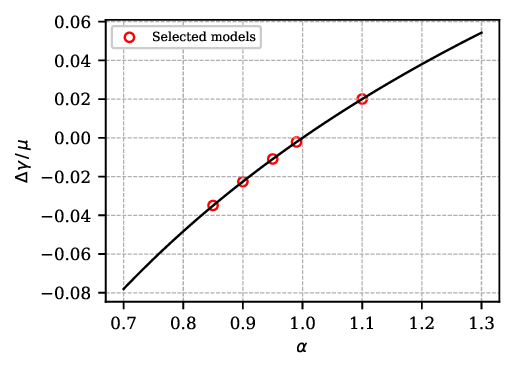}
\caption{\footnotesize The ratio $\Delta\gamma/\mu$ versus $\alpha$ (Eq.\,\ref{eq:DGamma}). Red circles indicate selected data points from Tab.\,\ref{tab1}.
}
\label{fig:D}
\end{figure}

\begin{figure} [t]
\centering
\includegraphics[width=87mm, trim={7pt 9pt 0pt 0pt}, clip]{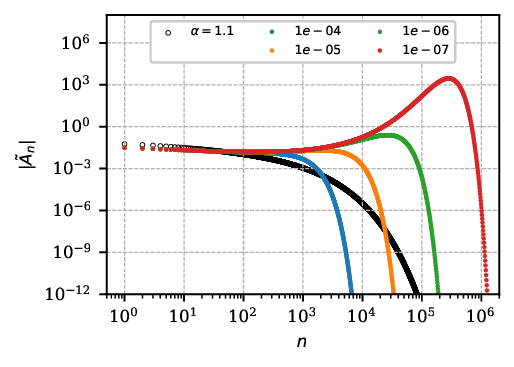}
\caption{Expansion coefficients $\tilde A_n$ of the least-damped eigenfunction $g(u)$ expanded over normalized Hermite polynomials for $n\ge 1$ ($\tilde A_0=\pi^{-1/4}$ in all cases). 
Colored circles (different colors indicate different $\mu$ values) depict $|\tilde A_n|$ for the stable case ($\alpha=0.95$), showing significant variations as $\mu \rightarrow 0$. 
Black open circles show $|\tilde A_n|$ for one unstable case ($\alpha=1.1$); here, coefficients remain nearly indistinguishable as $\mu$ decreases, converging to the collisionless limit ($\mu=0$).
}
\label{fig:A}
\end{figure}

\begin{figure*} [htbp]
\centering
\includegraphics[width=0.90\textwidth]{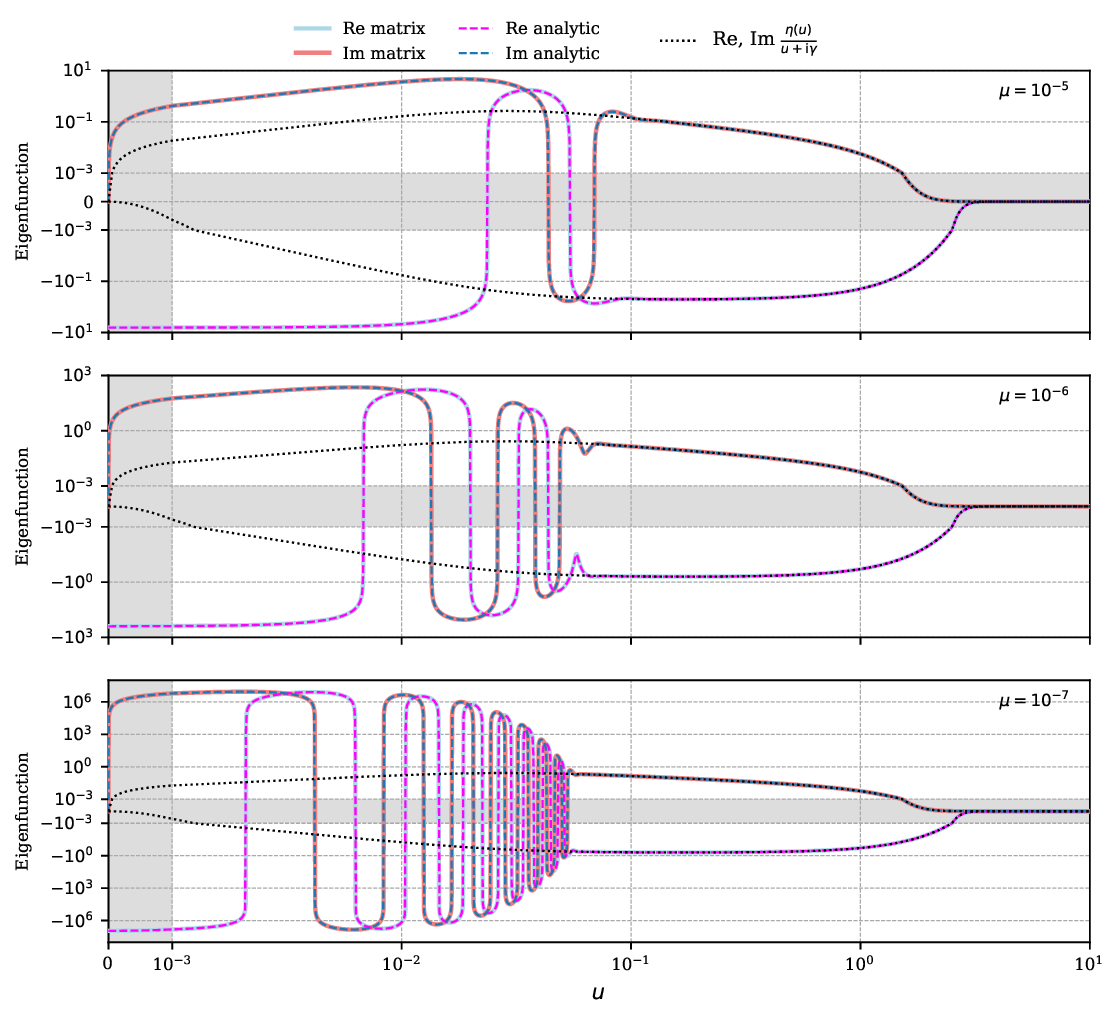}
\caption{\footnotesize  The normalized eigenfunction $g(u)$ of the least-damped mode is shown for $\alpha=0.95$ and $\mu=10^{-5}$, $10^{-6}$, and $10^{-7}$. Real parts of $g(u)$ are even functions, while imaginary parts are odd functions, and only the region $u\ge 0$ is displayed. Both the approximate analytic solution  (\ref{eq:hu}) and the exact numerical solution  (\ref{eq:gu}) are presented. Dotted lines represent the solution in the outer region (\ref{eq:g0}). As $\mu$ approaches zero, the number of oscillations and their amplitudes increase sharply. The solutions are plotted using symlog scaling for both coordinate axes. Symlog scaling combines logarithmic scaling with linear scaling near zero (indicated by grey shading).
}
\label{fig:EF-095}
\end{figure*}

\begin{figure*} [htbp]
\centering
\includegraphics[width=0.85\textwidth]{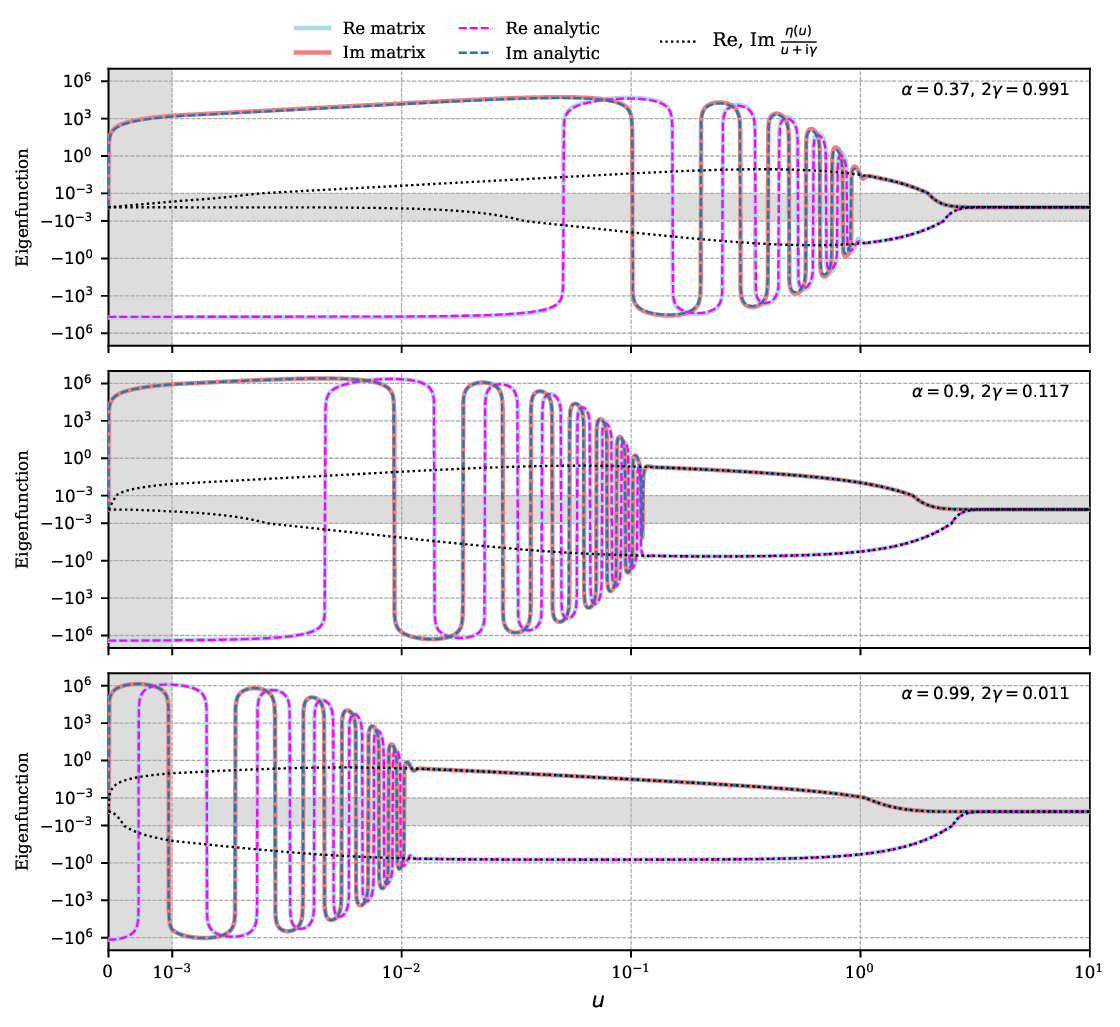}
\caption{\footnotesize  The normalized eigenfunction $g(u)$, analogous to Fig. \ref{fig:EF-095}, is shown for $\alpha=0.37$, $0.9$ and $0.99$ (see Tab.\,\ref{tab1}). These cases have approximate damping rate ratios of 0.5\,:\,0.05\,:\,0.005, respectively. The half-width of the resonance region, measured as the rightmost zero of the EF's imaginary part, is approximately $2\gamma$.
}
\label{fig:EF-b}
\end{figure*}

Our numerical validation of the analytical results is presented in Tab.\,\ref{tab1} and Figs.\,\ref{fig:D}--\ref{fig:EF-b}. Tab.\,\ref{tab1} shows calculated damping/growth rates $\gamma(\mu)$ for stable and unstable $\alpha$. The last column gives the numerical $\Delta\gamma/\mu = [\gamma(\mu)-\gamma_{\rm L}]/\mu$, comparable to the analytical $\gamma'$ from (\ref{eq:DGamma}), valid for small $\mu$ (last row of each block). High-precision calculations also allow us to infer the approximate second-order correction, valid for small $\mu$ and $|\beta|$ (figures not in table for brevity):
\be
    \gamma(\mu) \approx \gamma_{\rm L} -\frac{2}{3\pi}\,\beta \mu  - 0.014\, \beta \mu^2 \,,
\ee
where Eq.\,(\ref{eq:DGamma0}) has been taken into account. It is observed that the second-order correction acts in the same direction as the first-order correction, reducing the absolute value of the damping/growth rate and vanishing for the marginally stable case $\alpha=1$.

\begin{table}[t!]
\centering
\setlength{\tabcolsep}{0pt}
\begin{tabular}{>{\hspace{9pt}}c>{\hspace{9pt}}c>{\hspace{9pt}}c>{\hspace{-4pt}}c}

\toprule
$\alpha$ & $\mu$       & $\gamma$ & $\Delta\gamma/\mu$ \\ 
\midrule
\multirow[t]{5}{*}{0.37} & $0$         &$0.49552\;51860\;75220$\\[1mm]
\rowcolor{gray!20}       & $10^{-3}$   &&$-0.22789\;16095$ \\
                         & $10^{-4}$   &&$-0.22786\;68569$ \\
                         & $10^{-5}$   &&$-0.22786\;43821$ \\
&Formulae                &&$-0.22786\;41071$\\
\midrule
\multirow[t]{5}{*}{0.85} & $0$         &$0.08970\;84009\;02124$\\[1mm]
                         & $10^{-4}$   &&$-0.03498\;63258$\\
                         & $10^{-5}$   &&$-0.03498\;37029$\\
                         & $10^{-6}$   &&$-0.03498\;36791$\\ 
&Formulae                &&$-0.03498\;36764$\\
                         \midrule
\multirow[t]{5}{*}{0.90} & $0$         &$0.05860\;21552\;01507$\\[1mm]
                         & $10^{-5}$   &&$-0.02256\;80608$\\
\rowcolor{gray!20}       & $10^{-6}$   &&$-0.02256\;80460$\\
                         & $10^{-7}$   &&$-0.02256\;80445$\\  
&Formulae                &&$-0.02256\;80444$\\
\midrule
\multirow[t]{5}{*}{0.95} & $0$         &0.02873\;80319\;79385\\[1mm]
                         & $10^{-6}$   &&$-0.01093\;48149$\\ 
                         & $10^{-7}$   &&$-0.01093\;48142$\\ 
                         & $10^{-8}$   &&$-0.01093\;48142$\\ 
&Formulae                &&$-0.01093\;48141$\\
\midrule 
\multirow[t]{5}{*}{0.99} & $0$         &$0.00566\;25218\;19357$\\[1mm]
                         & $10^{-7}$   &&$-0.00213\;46775\;05$\\ 
                         & $10^{-8}$   &&$-0.00213\;46774\;92$\\ 
\rowcolor{gray!20}       & $10^{-9}$   &&$-0.00213\;46774\;90$\\ 
&Formulae                &              &$-0.00213\;46774\;90$\\
\midrule 
\multirow[t]{5}{*}{1.10} & $0$         &$-0.05448\;42519\;74364$\\[1mm]
                         & $10^{-5}$   &&$0.02004\;98460$\\ 
                         & $10^{-6}$   &&$0.02004\;98348$\\ 
                         & $10^{-7}$   &&$0.02004\;98337$\\ 
&Formulae                &&$0.02004\;98336$\\
\bottomrule
\end{tabular}

\vspace{2mm}
\caption{Damping rates $\gamma(\mu)$ and corrections $\Delta\gamma/\mu$ for the models shown. The first row in each block provides the Landau damping rate for $\mu=0$, $\gamma(0) = \gamma_{\rm L}$, with subsequent rows presenting corrections. 'Formulae' gives the analytic $\gamma'$ from Eq. (\ref{eq:DGamma}). The matrix size was $n_{\rm max}=3/\mu$, except for $(\alpha=0.99, \mu=10^{-9})$ where $n_{\rm max}=3\cdot10^{8}$. Models with $\alpha\leq 0.9$ were calculated using Advanpix (Matlab), and models with $\alpha>0.9$ using GMP (GNU). A negative damping rate for $\alpha=1.1$ indicates a growth rate. Corrections reduce the magnitude of both damping and growth rates. Grey rows denote models used in Fig.\,\ref{fig:EF-b}.}
\label{tab1}
\end{table}

Fig.\,\ref{fig:D} displays the curve of $\gamma'$, representing the first-order correction to $\gamma_{\rm L}$ as a function of $\alpha$, as given by Eq. (\ref{eq:DGamma}), along with numerical values of $\Delta\gamma/\mu$ for selected models from Tab.\,\ref{tab1}.

Fig.\,\ref{fig:A} illustrates the contrasting behavior of the eigenvalue vector ${\tilde A}_n$ in stable and unstable cases as $\mu \to 0$, approaching the collisionless limit. All components are real, with alternating signs from $n=1$. In the unstable case, the eigenvector components vary only slightly with $\mu$, making the markers nearly indistinguishable; therefore, we present only one eigenvector for $\alpha=1.1$ (black open circles). In contrast, for the stable case ($\alpha = 0.95$), the absolute values $|\tilde A_n|$ increase, reaching a maximum at $n\lesssim \gamma_{\rm L}/\mu$. This suggests a convergence issue with the sum in (\ref{eq:gu}) in the collisionless limit, reflecting the absence of eigenfunctions on the real $u$-axis in the purely collisionless case.

Figs.\,\ref{fig:EF-095} and \ref{fig:EF-b} compare numerical solutions for the normalized eigenfunctions $g(u)$ of the least-damped mode (solid lines) with the analytic expression (\ref{eq:hu}), shown as dashed colored lines. The zeroth-order approximation $g^{(0)}(u)$, given by (\ref{eq:g0}), is, in fact, the outer solution, and is shown as black dotted lines. The eigenfunctions exhibit symmetry: the real part is even, and the imaginary part is odd. Therefore, we present only the positive-$u$ portion. To accommodate the wide range of function variations, we use symlog scaling for both coordinate axes, combining logarithmic and linear scaling near zero, with linear regions indicated by grey shading. The final eigenfunction is calculated using (\ref{eq:gu}), truncating the sum at $n_{\rm fun}$, determined as the maximum $n$ for which $|A_n|$ exceeds $\varepsilon_{\rm tol}\cdot \max_n |\tilde A_n|$. By default, we used a Matlab routine with quadrupole precision and $\varepsilon_{\rm tol}=10^{-30}$ for $\alpha \leq 0.9$, and a C routine with higher precision (256 bits) and $\varepsilon_{\rm tol}=10^{-60}$ for more demanding calculations where $\alpha>0.9$.

Fig.\,\ref{fig:EF-095} shows three panels corresponding to gradually decreasing values of $\mu$ at fixed $\alpha=0.95$. In the outer region, the numerical solutions follow the zeroth-order approximation $g^{(0)}(u)$, but begin to oscillate as $u$ approaches zero. One can observe one oscillation for $\mu=10^{-5}$ in the top panel, two oscillations for $\mu=10^{-6}$ in the middle panel, and numerous oscillations for $\mu=10^{-7}$ in the bottom panel. This figure illustrates how the eigenfunction changes as $\mu$ decreases, with the evident absence of an eigenfunction in the collisionless limit.

Fig.\,\ref{fig:EF-b} illustrates the dependence of the resonance regions on the damping rates. We present three eigenfunctions $g(u)$ for values of $\alpha$ where the damping rates $\gamma$ are approximately in the ratio 0.5\,:\,0.05\,:\,0.005. The specific models used here are denoted by a grey background in Tab.\,\ref{tab1}. The half-width of the resonance region (i.e., for the positive part $u\geq 0$) is approximately $2\gamma$.

In all cases, the agreement between the numerical solutions obtained from (\ref{eq:gu}) and the approximate analytic expression given by (\ref{eq:hu}) is excellent.

\section{Conclusion}

This paper continues our study \cite{PSB} of an infinite homogeneous self-gravitating medium with a Maxwellian distribution function (DF), in which we present various specific properties of Landau damping solutions. In particular, we argued that these solutions are not true modes because an arbitrary perturbation of the DF, depending on real velocity $u$, does not decay self-similarly but becomes increasingly oscillatory in velocity space over time, despite the corresponding density perturbation decaying exponentially at sufficiently large times with the smallest Landau damping rate. The spectrum of eigenmodes for the stable Maxwellian DF is peculiar, consisting only of singular van Kampen modes \citep{vK_55, Case59}. It can be shown that Landau-damped solutions are, in fact, superpositions of van Kampen modes. On the other hand, we showed in \cite{PSB} that considering hypothetical perturbations on complex-valued $u$-contours allows presenting a true eigenfunction even for damped Landau solutions, but the contour must pass below the corresponding eigenvalue $c_{\rm L}$.

In a recent paper by \cite{Ng2021}, it was shown that introducing collisions eliminates van Kampen modes and transforms Landau quasi-modes into true eigenmodes. This is intuitively understandable, since even infrequent collisions suppress the formation of steep DF gradients in velocity space, from which van Kampen modes suffer. The authors presented numerical evidence that, as the dimensionless collision frequency $\mu$ tends to zero, the eigenvalue of the collision problem tends to the Landau-damped solution of the collisionless problem. Furthermore, in their earlier work \cite{Ng1999}, which focused on plasma, they presented numerical eigenfunctions for finite values of $\mu$.

In this paper, using our previous finding of the true eigenfunction on complex-valued $u$-contours, we derived the first-order correction to the eigenvalue with collisions. Notably, while the eigenfunctions become increasingly oscillatory and ultimately do not converge in the collisionless limit, the eigenvalues transition smoothly from the collisional to the collisionless regime. Our first-order correction term demonstrates this smooth eigenvalue transition. Then, using matched asymptotic expansions, we obtained approximate analytical solutions for the eigenfunctions, capturing their increasingly oscillatory behavior. We observe numerically that the resonance region's half-width is approximately $2\gamma$, and its amplitude grows without limit as $\mu$ approaches zero. It is also worth noting that although $2\gamma$ can be several times larger than $\delta$, this remains acceptable within the matched asymptotic expansion method, as these quantities are considered of the same order.

The diffusion form of the collisional term, specifically the presence of the second derivative, is crucial for the disappearance of van Kampen modes and the emergence of discrete eigenfunctions.  The simpler Krook collision term \citep[see ][] {BGK1954}, $-\nu\, \delta f(v,t)$, merely shifts the continious van Kampen spectrum of collisionless problem to the low half-plane $\omega$,  $\omega=- \i\nu+kv$,   retaining the absence of regular eigenfunctions. Conversely, the linearized Landau collision term \citep[see, e.g.,][]{lif_pit1995}, possessing a diffusion character, yields results qualitatively similar to those in \cite{Ng2021} and the present study.

The linearization of similar equations is a convenient and widely used method of analysis in plasma physics, with well-established applicability. Nonlinearity becomes significant when the damping rate $\gamma_{\rm L}$ is comparable to or smaller than the bounce frequency $\Omega_b$ of particles trapped in the potential wells of the wave. For our analysis to be valid, the gravitational potential amplitude must satisfy $\Phi\ll (\gamma_{\rm L}/k)^2$. For scenarios violating this inequality, we refer the reader to \citet{ONeil1965} and \citet{Gal_Sag1979} for a description of perturbation evolution in the essentially nonlinear regime.

Numerical validation confirmed our analytical findings, including the adopted ordering of scales. In contrast to \cite{Ng1999}, we employed standard matrix eigenvalue problem solvers, which are particularly efficient for sparse tridiagonal matrices. Using a moderate workstation with 256 GB of memory, we are able to handle matrices with $n_{\rm max} = 10^9$ with high precision accuracy. Given that the maximum of the eigenvalue coefficients $|\tilde A_n|$ is attained at $n\lesssim \gamma/\mu$, and that several multiples of this $n$ are required for convergence of the eigenvalue and eigenvector calculation methods, we can achieve values of $\mu$ on the order of $10^{-9}\beta$. Here, $\beta=1-\alpha$ is a parameter that characterizes the proximity to the stability boundary of the Maxwellian DF.

The code is freely available at the GitLab repository: \url{https://gitlab.com/epolyach/collision_gravity_rare_collisions}.

\begin{acknowledgments}
We thank M. Weinberg for pointing out the papers by Ng, Bhattacharjee \& Skiff and the anonymous referee for his valuable comments and suggestions. This research was partially supported by NSF grant PHY-2309135 to the Kavli Institute for Theoretical Physics (E. Polyachenko), and by the Russian Academy of Sciences Program No. 28 (subprogram II, 'Astrophysical Objects as Cosmic Laboratories) and the Ministry of Science and Higher Education of the Russian Federation (I. Shukhman). High-precision calculations were performed with the Advanpix Multi-precision Computing Toolbox and the GNU Multiple Precision Floating-Point Reliable Library. 

\end{acknowledgments}

\appendix
\section{Evaluation of the Integrals $I_{1,2}$}
For $I_1$, we have
\be
I_1\equiv \frac{\alpha}{\sqrt{\pi}}\int\limits_\curvearrowbotright\hspace{-2pt}\d u\, \frac {u\,\e^{-u^2}}
{(u-c_{\rm L})^2} =
\frac{\alpha}{\sqrt{\pi}}
\int\limits_\curvearrowbotright \hspace{-2pt}\d u \, \frac{\e^{-u^2}}{u-c_{\rm L}}+
c_{\rm L}\frac{\alpha}{\sqrt{\pi}}\int\limits_\curvearrowbotright \hspace{-2pt}\d u \, \frac{\e^{-u^2}}{(u-c_{\rm L})^2} = J-2 c_{\rm L}\frac{\alpha}{\sqrt{\pi}}\int\limits_\curvearrowbotright \hspace{-2pt}\d u \, \frac{u\,\e^{-u^2}}{u-c_{\rm L}} = J-2c_{\rm L}\,,
\ee
where, for the last equality, Eq. (\ref{eq:Deq2}) was used, and we introduced
\be
J \equiv \frac{\alpha}{\sqrt{\pi}}\int\limits_\curvearrowbotright \hspace{-2pt}\d u \, \frac{\e^{-u^2}}{u-c_{\rm L}}.
\ee
The integral $J$ can be readily calculated from the dispersion equation.
We have
\be
1=\frac{\alpha}{\sqrt{\pi}}
\int\limits_\curvearrowbotright \hspace{-2pt}\d u\,
\frac {u \e^{-u^2}}
{u-c_{\rm L}}=
\frac{\alpha}{\sqrt{\pi}}\int\limits_{-\infty}^{\infty} \hspace{-2pt}\d u\,
\e^{-u^2} +c_{\rm L}J=\alpha +c_{\rm L}J
\ee
Hence $J = (1-\alpha)/{c_{\rm L}} = \beta/{c_{\rm L}}$ and, finally we obtain
\be
I_1=\frac{\beta-2c_{\rm L}^2}{c_{\rm L}}.
\ee

For $I_2$, after three integrations by parts,
\begin{multline}
I_2= \frac{\alpha}{\sqrt{\pi}}
\int\limits_\curvearrowbotright \hspace{-2pt}\d u
\frac{\e^{-u^2}}{(u-c_{\rm L})^4}=
\frac{2}{3}\frac{\alpha}{\sqrt{\pi}}
\int\limits_\curvearrowbotright \hspace{-2pt}\d u
\frac{\e^{-u^2}}{u-c_{\rm L}} (3u-2u^3) =\frac{2}{3}\frac{\alpha}{\sqrt{\pi}}
\int\limits_\curvearrowbotright \hspace{-2pt}\d u 
\frac{\e^{-u^2}}{u-c_{\rm L}}\Bigl[-(u-c_{\rm L})^3+3(u-c_{\rm L})^2 c_{\rm L} \\
+3(u-c_{\rm L})c_{\rm L}^2+c_{\rm L}^3\Bigr] 
=\frac{2}{3}\frac{\alpha}{\sqrt{\pi}}
\int\limits_{-\infty}^{\infty} \hspace{-2pt}\d u\,
\e^{-u^2}\Bigl[-(u-c_{\rm L})^2+3\,(u-c_{\rm L})c_{\rm L}+3\,c_{\rm L}^2\Bigr] -\frac{2}{3}c_{\rm L}^3 J \\
=\frac{2}{3}\frac{\alpha}{\sqrt{\pi}}
\int\limits_{-\infty}^{\infty} \hspace{-2pt}\d u\,
\e^{-u^2}(u^2+c_{\rm L}^2)  - \frac{2}{3}\,c_{\rm L}^3 J =\frac{2}{3}\alpha+\frac{4}{3}\,\alpha\, c_{\rm L}^2-\frac{2}{3}\,c_{\rm L}^2(1-\alpha)
=\frac{4}{3}+\frac{2}{3}\left(\beta-2 c_{\rm L}^2\right).
\end{multline}
Finally,
\be
I_2=\frac{4}{3}+\frac{2}{3}\left(\beta-2 c_{\rm L}^2\right)=\frac{4}{3}+\frac{2}{3}\,c_{\rm L}\,I_1.
\ee


\bibliography{main}{}
\bibliographystyle{aasjournal}

\end{document}